\documentclass{appolb}
\usepackage{epsfig}
\usepackage{graphicx}
\usepackage{amsmath}
\usepackage{caption}
\usepackage{color}
\definecolor{gray}{gray}{0.4}

\usepackage{amssymb}

\usepackage{makeidx}

\def\simge{\mathrel{%
    \rlap{\raise 0.511ex \hbox{$>$}}{\lower 0.511ex \hbox{$\sim$}}}}
\def\simle{\mathrel{
    \rlap{\raise 0.511ex \hbox{$<$}}{\lower 0.511ex \hbox{$\sim$}}}}

\newcommand \be{\begin{eqnarray}}
\newcommand \ee{\end{eqnarray}}

\def\XXint#1#2#3{{\setbox0=\hbox{$#1{#2#3}{\int}$}
\vcenter{\hbox{$#2#3$}}\kern-.5\wd0}}

\begin{document}

\title{Collective Correlations of Brodmann Areas \\ fMRI Study with RMT-Denoising}

\author{
Zdzislaw Burda$^{1,2}$, 
Jennifer Kornelsen$^3$, 
Maciej A. Nowak$^{1,2}$, 
Bartosz Porebski$^1$, 
Uta Sboto-Frankenstein$^4$, \\
Boguslaw Tomanek$^5$,
Jacek Tyburczyk$^1$\thanks{jacek.tyburczyk@uj.edu.pl}
\address{1: M. Smoluchowski Institute of Physics, Jagiellonian University, PL--30--059 Cracow, Poland \\ 
2: Mark Kac Center for Complex Systems Research, Jagiellonian University, PL--30--059 Cracow, Poland \\
3: University of Winnipeg, Winnipeg, Manitoba, Canada \\
4: Alberta Innovates, Winnipeg, Manitoba, Canada \\
5: Thunder Bay Regional Research Institute, Thunder Bay, Ontario, Canada
Institute of Nuclear Physics, Polish Academy of Sciences, Krakow, Poland
Alberta Innovates - Technology Futures, Calgary, Alberta, Canada
}}

\maketitle

\begin{abstract}
We study collective behavior of Brodmann regions of human cerebral cortex using 
functional Magnetic Resonance Imaging (fMRI) and Random Matrix Theory (RMT).
The raw fMRI data is mapped onto the cortex regions corresponding to the Brodmann areas with the aid of the Talairach coordinates. Principal Component Analysis (PCA) of the Pearson correlation matrix for 41 different Brodmann regions is carried out to determine their collective activity in the idle state and in the active state stimulated by tapping. The collective brain activity is identified through the statistical analysis of the eigenvectors to the largest eigenvalues of the Pearson correlation matrix. The leading eigenvectors have a large participation ratio. This indicates that several Broadmann regions collectively give rise to the brain activity associated with these eigenvectors. We apply random matrix theory to interpret the underlying multivariate data.

\end{abstract}
\PACS{PACS numbers come here}
\section{Introduction}
The problem of understanding brain activity is of fundamental importance in basic research and clinical neuroscience. In recent years several advanced physical techniques, such as electroencephalography (EEG), electrocorticography (ECoG) or magnetoencephalography (MEG), and the functional Magnetic Resonance Imaging (fMRI), have been developed to monitor the brain activity. The latter method uses the hemodynamic response related to the oxygenation level of blood. 

A common feature of these methods is that they require a massive accumulation and analysis of data, which is usually contaminated by statistical noise. In order to extract the relevant information, several algorithms for maximizing the signal to noise ratio and for localization of the sources of specific neuronal activities have been developed.
Due to the high-dimensionality of the system in question, its complex nature, nonlinearity, potential non-stationarity and emerging collective behavior, problem is becoming hard to solve using the traditional methods of multivariate statistical analysis. In many respects it resembles probabilistic problems encountered in different areas of research and engineering including contemporary wireless networks~\cite{TULINO-VERDU,DEBBAH}, financial markets~\cite{BOUCHAUD,BURDAJAROSZNOWAK} and complex biological systems~\cite{ZEEORLAND,GUDOWSKA,SHNERB}, to name few of them, where one deals with huge multivariate data sets. Recently in order to come forward  researchers in these fields have started to borrow ideas from emergent domains of physics and mathematics such as statistical theory of networks, percolation theory, spin glasses, random matrix theory, free random probability, game theories, etc. We believe that the same methodology can be applied to neuroscience. In this letter we start this program by applying random matrix theory to address the problem of the brain activity detection in large multivariate experimental data sets. 

More specifically, we use fMRI data to study collective correlations of Brodmann areas. Those are specific regions of brain with empirically assigned role. More details can be found in the section \ref{s_DS} and table \ref{brodmann_numbers} below. In our experimental setup the brain is stimulated by some simple tasks. We measure the activity of different Brodmann areas and try to determine the collective behavior of several areas under a given stimulus. We focus not on the response mechanism to a particular stimulus of an individual brain area but rather on the collective response of many Brodmann areas. We apply principal component analysis to the fMRI data to extract eigenspaces of the matrix of the Pearson coefficients for pairs of fMRI signals coming from different Brodmann areas. The largest eigenvectors of this matrix give collections of Brodmann areas that take part in collective independent brain activities. 

We supplement this analysis by showing how to apply random matrix theory in multivariate analysis of the experimental data to maximize the signal to noise ratio. The random matrix methodology is general and it can be applied to analyze other multivariate data sets in neuroscience like those in EEG and MEG. Actually, random matrix theory has already been introduced in the analysis of neurophysiological data. More than a decade ago, Dro\.{z}d\.{z} et al \cite{DROZDZ} used the nonhermitian random matrix model for analyzing the MEG data. In 2003, Seba~\cite{SEBA} suggested to use random matrix theories in order to identify generic and subject-independent features of certain correlation matrices extracted from human EEG. Then Beckmann and Smith~\cite{BS} have proposed to modify the Independent Component Analysis (ICA) algorithm commonly used for fMRI data in order to take into account the asymmetry between the number of observations and the number of sources and the mixing of the noise, constituting the probabilistic ICA models (PICA), resembling the MIMO (multiple input multiple output) systems based on Wishart ensemble. Only very recently, the random matrix theory (RMT) approach has been suggested as a tool to analyze temporal correlations in EEG data in order to find synchronization patterns characteristic for seizures in epileptic attacks~\cite{OSORIO-LSAI}. 

The paper is organized as follows. In Section \ref{s_RMT} we outline the basic result of RMT on multivariate data and sketch some general ideas on large random matrices and free probability. 
Then we  recall how to derive the well known Marchenko-Pastur distribution~\cite{MARCENKOPASTUR}. This law serves as a calibrating tool in multivariate statistical analysis providing one with a benchmark for uncorrelated data. Deviations between this benchmark and spectral properties of the real data signal potential correlations. In Section \ref{s_exp} we describe the experimental set-up of fMRI experiments. In Section \ref{s_DS} we recall Talairach coordinates,  describe the data structure and provide some basic information on physiological activities of the Brodmann areas. In Section \ref{s_R} we present the results. We compare the data to the Marchenko-Pastur benchmark, identify outliers and discuss their interpretation.  
We conclude the paper in Section \ref{s_C}, were we shortly summarize the paper and discuss generalizations and perspectives.

\section{Large matrices and freeness}\label{s_RMT}

Random matrices find nowadays ubiquitous applications in many branches of science. The reason for this is two-fold. First, random matrices posses a great degree of universality that is: eigenvalue properties of large matrices do not depend on details of the underlying statistical matrix ensemble. Second, random matrices can be viewed as non-commuting random variables. As such they form a basis of a non-commutative probability theory where the whole matrix is treated as an element of the probabilistic space. In the limit when the size of the matrix tends to infinity, the connection to the probability theory is becoming exact in the mathematical sense. This is the celebrated free probability theory, where independent matrices play the role of free random variables (hereafter FRV)~\cite{VOICULESCU}. Nowadays, data sets are usually organized as large matrices whose first dimension is equal to the number of degrees of freedom and the second to the number of measurements. Typical examples are large economic/financial systems, wireless networks and genetic data. In all these fields, FRV found already important applications. It is therefore tempting to challenge the power of free random variables on  human brain data, where the recorded files can easily take several gigabytes per person per session. In this work we study the simplest case of so-called free Poisson process, that leads to a well known Marchenko-Pastur~\cite{MARCENKOPASTUR} spectral distribution. We prefer to look at this process from the perspective of FRV, which allows one for an easy incorporation of time and/or space correlations into the data \cite{BURDAJAROSZNOWAK}. 
The eigenvalue density of the empirical correlation matrix for uncorrelated independent identically distributed (i.i.d.)
Gaussian variables are given in the large matrix size limit by the following distribution:
\be
\rho(\lambda)=\frac{1}{2\pi r \lambda}  \sqrt{(\lambda_+-\lambda)(\lambda-\lambda_-)}
\label{mp}
\ee
which is usually called Marchenko-Pastur distribution or Wishart distribution.
Here the rectangularity parameter $r=N/T$ is the aspect ratio between number of variables and the the sample size in the data matrix.
According to this formula eigenvalues of the correlation matrix are
located in a finite support $\lambda \in [\lambda_- ,\lambda_+]$ whose
end-points are $\lambda_\pm = (1 \pm \sqrt{r})^2$. This result plays a similar role for multivariate analysis as normal distribution for univariate statistics. One can see that, when $r\rightarrow 0$ ($T$ is much larger than $N$) then $\rho(\lambda)$ approaches a delta function, exactly as one expects for genuine correlation matrix of uncorrelated i.i.d. variables. However when $r=N/T$ is finite the eigenvalue density is smeared. The case $N=T$ $(r=1)$ is critical, since then $\lambda_-=0$. When the number of measurements $T$ is less than the number of degrees of freedom $N$ one is unable to determine the underlying correlation function and that is signaled by the appearance of $T-N$ zero  eigenvalues in the spectrum. We shall not discuss this case here. The distribution (\ref{mp}) serves us as a reference point - a benchmark to look for non-trivial correlations between degrees of freedom in the data sets. 

\section{Experiment} \label{s_exp}

Short description of the experiment itself is presented in this section.
\\
\\
\textbf{Subject}:\\
The subject was a 61 year old healthy male, right-handed and performed the motor task with the right hand. 
\\
\\
\textbf{MR Data Acquisition:}\\
Data was collected at the National Research Council of Canada Institute for Biodiagnostics MR facilities in Winnipeg and approved by the NRC`s Human Research Ethics board.  Informed consent was obtained prior to the subject's entry into the magnet. Experiments were conducted with a $3$ Tesla Siemens TRIO whole body magnet with a homogeneous birdcage coil. Conventional BOLD imaging techniques were used. Whole brain EPI images were referenced and acquired parallel to the AC-PC line (anterior/posterior commissure, \cite{TALAIRACH}). Single-shot blipped gradient-echo planar images were acquired with the following parameters: TR/TE $= 3000/60~msec$, flipangle $= 80^{\circ}$, $64 x 64$ matrix, $25~cm$ FOV.  High-resolution $T1$-weighted gradient-echo images will be obtained for the overlay of functional activation maps.  Whole brain axial slices were acquired using a spoiled gradient-echo sequence ($1.5~mm$ slice thickness, field of view $= 25~cm$, in plane resolution $0.94 \times 0.94~mm$, TE $=5~ms$, TR $=24~ms$).  
During the motor task, the participant was instructed to observe a display which indicated either ''REST'' or ''TAP FINGERS''. During the 'rest' display the subject was instructed to simply lay still and rest. During the "tap fingers" display the subject performed an alternating finger tapping sequence of thumb to index, ring, middle, and pinky finger repeatedly. The participant was instructed to pay attention to this finger tapping task and to not perform it absentmindedly. The motor task paradigm consisted of four one-minute blocks interspersed with one minute rest blocks, for a total of $8$ minutes during which $163$ volumes were acquired.  The subject started to tap his fingers on volume $1$ and continued in the following sequence: volumes $1-22$, $43-62$, $83-102$, $123-142$.

\section{Data Structure} \label{s_DS}

Brain sizes and shapes vary greatly among individuals. This makes a direct comparison of spatial data very difficult. Furthermore, even consecutive surveys of the brain activity of the same person are prone to a translation and rotation of the subject, resulting in sets of completely different slices.

In order to enable spatial comparisons between different scans and different individuals a standardization is required that brings different measurements to a common reference frame with the given coordinate system. A commonly used coordinate system is a Talairach space - a reference frame proposed by Jean Talairach in his atlas of anatomy \cite{TALAIRACH}. We use it throughout this paper. 

The most important landmark for the Talairach space is the anterior commissure. It is a fibre tract connecting the two hemispheres, running just in front of the fornix. It is the origin of the Talairach space. The coordinates are written in format (x,y,z), where the X axis refers to left-right, Y to posterior-anterior, and Z to ventral-dorsal.

To transform a dataset onto the Talairach space two transformations are required.
The first one is translation and rotation of the coordinate system. Two points need to be fixed: the first is the aforementioned anterior commissure and the second one is the posterior commissure. The line between them becomes the horizontal Y axis. Two other axes are defined by the plane separating the hemispheres. In the next step
the scan is warped onto standardized dimensions of the brain. This transformation is not linear and takes into account both commissures as well as the farthest points of the cerebrum in all directions. The standardized dimensions are: length - 172 mm, height - 116 mm, width - 136 mm. Jean Talairach based his atlas on postmortem sections of a subject with a less than average brain size, therefore in most cases the scans are shrunk in this step.

As mentioned in the introduction, fMRI techniques are capable of measuring brain activity, through hemodynamic response, at the scale of millimeters. The most common approach in the data analysis is to use purely geometrical voxels and to neglect all the information about cellular organization and brain inner structure. In the present work we use Brodmann areas as elementary units for which we want to study correlations. The cerebral cortex of a brain is divided into 52 Brodmann areas. The division is based on cytoarchitectonics.  
For over century now the Brodmann classification has been the subject of many intensive studies and still remains a most widely known and frequently cited brain atlas. Introduction of such a division for the data allows us to analyze correlations, taking into account the physiological aspects and to compare our results with those known from previous neurological studies. 

In practice each Brodmann area contains many voxels, so in the first of the analysis we ascribe voxels to Broadmann areas by comparing voxels' spatial coordinates with Brodmann regions boundaries. Then, for each time slice and for each Brodmann area we calculate the mean value of fMRI signals for all voxels in the given region. In this way we obtain $N$ time series, where $N$ is the number of analyzed Brodmann areas.
These times series form the basis for further analysis. The data points $y_{it}$ are indexed by an index $i=1,\ldots,N$ that runs over the set of Brodmann areas and $t=1,\ldots,T$ that runs over the set of $T$ consecutive measurements at different times.
One may ask whether the arithmetic mean should be used in this case, since regions are not perfectly fitted. An optional approach would be to calculate a weighted mean with a weight that varies with the voxel position within the given Brodmann area. In this prescription voxels nearer to the region center would contribute more than those on boundaries. This would be however more complicated and there would be still some ambiguity in assigning the weights. We know, for example, that neurons responsible for performing some tasks (e.g processing visual stimulus) are located in one Brodmann region, but this distribution is not uniform. Different aspects of processing may cause activation in different parts of the region. So we see, that calculation of the weighted mean might enhance some uncontrolled effects and thus lead to larger uncertainties. To summarize, the simplest choice corresponding to the arithmetic mean seems to be the easiest and for the moment the best for our purposes.

\section{Results} \label{s_R}

We calculate the Pearson correlation matrix for Brodmann areas. 
Given a time series $y_{it}$ for $i$-th Brodman region, $t=1,\ldots, T$,
we first standardize it to obtain a corresponding time series $x_{it}$ with 
zero mean and unit variance:
\be
x_{it} = \frac{1}{\sigma_i}\left(y_{it}-\bar{y}_i\right) \label{stand} 
\ee
where $\bar{y}_i$ is the estimated mean $\bar{y}_i = \frac{1}{T}\sum_{t=1}^T y_{it}$
and $\sigma_i$ the standard deviation of the original
time series, that is estimated by  $\sigma_i^2= \frac{1}{T-1} \sum_{t=1}^T (y_{it} - \bar{y}_{i})^2$. Such a rescaling is quite standard and removes a potential 
heterogeneity that might result in magnifying the importance of certain time series while reducing the contribution of others. 
We can calculate Pearson correlation matrix using the standardized time series 
\be 
P_{ij} = \frac{1}{T-1}\sum_{t=1}^T \frac{\left(y_{it}-\bar{y}_i\right) \left(y_{jt}-\bar{y}_j\right)}{\sigma_i \sigma_j} 
= \frac{1}{T-1}\sum_{t=1}^T x_{it} x_{jt}
\label{pc}
\ee
where in this particular case the indices $i$ and $j$ run over the set of $N=41$ Brodmann areas. The original atlas consists of 52 regions, but a few are too small for fMRI resolution. For computational convenience we have changed the numbering of areas. The correspondence between the original atlas and our numbering is given in Table \ref{brodmann_numbers}.
\begin{table}[t]
\begin{minipage}[t]{.7\textwidth}
	\centering
	\begin{tabular}{ | c | c | c | }
\hline 
Area No. & Ref. To & Description \\ 
\hline 
1 & 1 & Primary Somatosensory C. \\
2 & 2 & Primary Somatosensory C. \\
3 & 3 & Primary Somatosensory C. \\
4 & 4 & Primary Motor C. \\
5 & 5 & Somatosensory Association C. \\
6 & 6 & Premotor C. \\
7 & 7 & Somatosensory Association C. \\
8 & 8 & Part of the Frontal C.\\
9 & 9 & Dorsolateral Prefrontal C. \\
10 & 10 & Anterior Prefrontal C. \\
11 & 11 & Orbitofrontal Area \\
12 & 17 & Primary Visual Cortex \\
13 & 18 & Secondary Visual Cortex \\
14 & 19 & Associative Visual Cortex \\
15 & 20 & Inferior Temporal Gyrus \\
16 & 21 & Middle Temporal Gyrus \\
17 & 22 & Superior Temporal Gyrus \\
18 & 23 & Posterior Cingulate C. \\
19 & 24 & Anterior Cingulate C. \\
20 & 25 & Subgenual C. \\
21 & 26 & Ectosplenial area 26\\
22 & 27 & Piriform C. \\
23 & 28 & Posterior Entorhinal C. \\
24 & 29 & Retrosplenial Cingulate C. \\
25 & 30 & part of Cingulate C. \\
26 & 32 & Anterior Cingulate C. \\
27 & 34 & Anterior Entorhinal C. \\
28 & 35 & Perirhinal C. \\
29 & 36 & Parahippocampal C. \\
30 & 37 & Fusiform Gyrus \\
31 & 38 & Temporopolar Area \\
32 & 39 & Angular Gyrus \\
33 & 40 & Supramarginal Gyrus \\
34 & 41 & Primary Association C. \\
35 & 42 & Auditory Association C. \\
36 & 43 & Primary Gustatory C. \\
37 & 44 & Pars Opercularis \\
38 & 45 & Pars Triangularis \\
39 & 46 & Dorsolateral Prefrontal C. \\
40 & 47 & Inferior Prefontal Gyrus \\
41 & 48 & Retrosubicular Area \\
\hline
	\end{tabular}
	\captionof{table}{Indexing of Brodmann areas used in our analysis (first column) and its reference to the standard atlas.}
	\label{brodmann_numbers}
\end{minipage}\qquad
\hfill
\begin{minipage}[t]{.23\textwidth}
	\centering
	\begin{tabular}{ | c | c | }
\hline
$eigv_1$ & $eigv_2$ \\ 
\hline 
0.099 & 0.31 \\
0.17 & 0.15 \\
0.17 & 0.19 \\
0.18 & 0.18 \\
0.16 & 0.21 \\
0.17 & 0.22 \\
0.17 & 0.2 \\
0.17 & 0.042 \\
0.18 & 0.063 \\
0.07 & -0.29 \\
0.082 & -0.31 \\
0.17 & 0.063 \\
0.19 & 0.055 \\
0.18 & 0.079 \\
0.17 & -0.13 \\
0.18 & -0.091 \\
0.18 & -0.069 \\
0.19 & -0.059 \\
0.18 & 0.058 \\
0.061 & -0.33 \\
0.11 & -0.069 \\
0.12 & -0.22 \\
0.027 & -0.17 \\
0.1 & -0.14 \\
0.15 & -0.14 \\
0.17 & 0.071 \\
0.11 & -0.042 \\
0.076 & -0.036 \\
0.12 & -0.16 \\
0.19 & -0.095 \\
0.14 & -0.25 \\
0.19 & 0.064 \\
0.19 & 0.12 \\
0.15 & -0.067 \\
0.17 & -0.067 \\
0.16 & 0.057 \\
0.19 & 0.011 \\
0.19 & 0.038 \\
0.17 & -0.052 \\
0.13 & -0.28 \\
0.2 & -0.042 \\
\hline
	\end{tabular}
	\captionof{table}{Eigenvectors to the first and second largest eigenvalues of the Pearson correlation matrix.}
	\label{eigv_t}
\end{minipage}
\end{table}
The Pearson correlation matrix is symmetric and positive semidefinite. Moreover it has unities on the diagonal, so the trace of this matrix is equal $N$. Values of the matrix elements $P_{ij}$ lie in the range $\left[-1;1\right]$ and are interpreted as correlation coefficients between $i$-th and $j$-th Brodmann regions. Such a matrix can be diagonalized. Due to the trace invariance the sum of its eigenvalues is equal $N$.

\begin{figure}[ht]
	\centering
	\includegraphics[angle=0,scale=0.30]{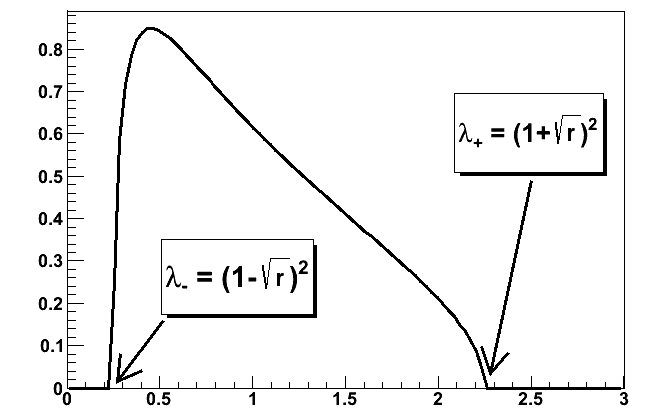}
		\caption{Marchenko-Pastur distribution for $r=0.256$.}
		\label{mpfig}
\end{figure}

While individual entries of the Pearson correlation matrix tell us about the mutual correlations of pairs of Brodmann regions, eigenvalues and eigenvectors of this matrix bear a very interesting information about the collective activity of the whole brain in which simultaneously participate many Brodmann areas. Such a collective behvior is signalized in principal component analysis by the appearance of large eigenvalues sticking out from the bulk of the eigenvalue spectrum given by the Marchenko-Pastur benchmark. As we have discussed before, if the signals coming from the Brodmann regions were completely uncorrelated, the density of eigenvalues would form the Marchenko-Pastur distribution with the support $[\lambda_-,\lambda_+]$ (\ref{mp}). In our case the end points of the spectrum would be located at $\lambda_-\approx 0.25, $ $\lambda_+\approx 2.25$ since the aspect ratio $r=N/T\approx 0.256$. The corresponding probability density function with such parameters is shown in Figure~\ref{mpfig}.
Comparing the pdf and eigenvalues summarized in Table \ref{t_eigenvalues} we see that at least two of them are outliers and therefore must be related to actual collective correlations of Brodmann areas. In next subsections we discuss the parameter $\lambda_+$ and the two largest eigenvalues in more detail.

\begin{table}
\centering
\begin{tabular}{|c|c|}
\hline
n & n-th eigenvalue \\
\hline
1 & 22.627\\
2 & 4.478\\
3 & 2.113\\
4 & 1.623\\
5 & 1.420\\
6 & 1.211\\
7 & 1.014\\
8 & 0.781\\
9 & 0.727\\
10 & 0.546\\
\hline
\end{tabular}
\caption{10 largest eigenvalues of the empirical correlation matrix and the
inverse participation ratios of the corresponding eigenvectors.}
\label{t_eigenvalues}
\end{table}

\subsection{$\lambda_+$ as a cutoff criterion}
The main idea behind principal component analysis is to find an orthogonal transformation that transforms a multivariate system of correlated random variables to new coordinates that are linearly uncorrelated. They are called principal components and in the case of normal distributions they represent independent constituents (features) of the underlying statistical system. The task is formally accomplished by a decomposition of the Pearson correlation matrix $P = U^T \Lambda U$, where $U$ is an orthogonal one and $\Lambda$ - a positive semidefinite diagonal matrix. Such a transformation exists since the matrix $P$ is by construction symmetric and positive semidefinite. Using the orthogonal transformation we can construct  time series corresponding to principal components by calculating the weighted mean of initial data:
\begin{equation}
\label{PCA_eq}
\bar{x}_a(t) = \sum_{i=1}^N x_i(t) U_{i,a}
\end{equation}
From now on, we will refer to them as \textit{eigenseries}. It is easy to show that $cov(\bar{x}_a,\bar{x}_b)=\delta_{a,b}$ and $var(\bar{x}_a) = \lambda_a$. 

The transformation is revertible because the matrix $U$ is orthogonal. The original time series can be recalculated using following formula:
\begin{equation}\label{f_inv}
x_i(t) = \sum_{a=1}^N \bar{x}_a(t) U_{a,i} \ .
\end{equation}
One can see, according to the formula (\ref{f_inv}), that the larger the variance of eigenseries is, the larger is its contribution to the original time series. The variance of the eigenseries is equal to the corresponding eigenvalue of the correlation matrix. The main idea behind PCA is to reduce the complexity of the problem by neglecting a part of less significant degrees of freedom. It is clear that one should first neglect those eigenseries that have smallest variances. In practice, one often reduces the number of principal components to only a few leading ones. It is an  open question  what is the optimal choice of the number of significant principal components. In the literature on the subject the choice varies from study to study and the criteria are rather heuristic. For example one sometimes assumes that one should take as many eigenvalues as to keep their net contribution to the spectrum between 70\% and 90\%, but the exact number depends on the desired level of the balance between the confidence level and  the feature space dimension. 

Such a criterion does not take into account the proportion between number of features and the number of observations and it contradicts the intuition that the number of nonsignificant principal components should depend on the signal-to-noise ratio. The gap can be filled up by a RMT analysis.  
According to this analysis random eigenvalues are described by the Marchenko-Pastur distribution localized on an interval $[\lambda_-,\lambda_+]$ (that is sometimes called the bulk of the distribution). Eigenvalues that carry a non-statistical information typically lie outside this interval.  The end-points of the interval depend on the aspect ration $N/T$ that is related to the signal-to-noise ratio. Most of the PCA criteria neglect the dependence on $N/T$. We propose a new criterion that directly refers to $\lambda_+$ and thus also to $N/T$. The main idea behind this criterion is that eigenvalues smaller than $\lambda_+$ can occur just as a result of a finite sample size. In the new criterion we treat $\lambda_+$ as a cut-off value to select the eigenvalues corresponding to significant principal components for which the signal exceeds the statistical noise.

\subsection{Eigenseries corresponding to $\lambda_1$}
The largest eigenvalue of the Pearson correlation matrix for the analyzed dataset has a magnitude of $22.627$ which is more than a half of the whole spectrum. According to the Marchenko-Pastur criterion formulated in the previous subsection, it is greater than $\lambda_+$ and therefore the information content of the corresponding eigenvector exceeds the statistical noise. The elements of this eigenvector calculated for our data are shown in Table \ref{eigv_t}. As one can see the estimated values of the elements are of the same order. This means that all Brodmann regions give rise to this eigenvector, or in other words,  the corresponding brain activity involves all regions. This result can be quantified by measuring the inverse participation ratio for this eigenvector, the concept borrowed from the localization theory~\cite{ANDERSON}. Actually for our purposes it is more convenient to use its inverse that is the participation ratio being a statistical measure of the number of non-zero vector components. In quantum systems, participation ratio $PR$ is defined as $PR=1/(\sum_i p_i^2)$, where $p_i$ is the probability that the particle is in the state $i$, given by the modulus squared of the wave function. 
Since in our case the role of the "wave functions"is played by the eigenvectors $v_i$, we define the  the participation ratio as
$$PR(v) = \frac{1}{\sum_{i=1}^N v_i^4} \ .$$
In the limiting case when all but one elements of the eigenvector are zero $PR=1$. Such eigenvectors are called fully localized. The other extreme case is when all elements have the same value. In this case gives $PR=N$. In this case the  eigenvectors are called fully delocalized. In the intermediate cases when for example $N-n$ elements are zero and $n$ have the same non-zero value $PR=n$. 

\begin{figure}
	\centering
	\includegraphics[angle=0,scale=0.3]{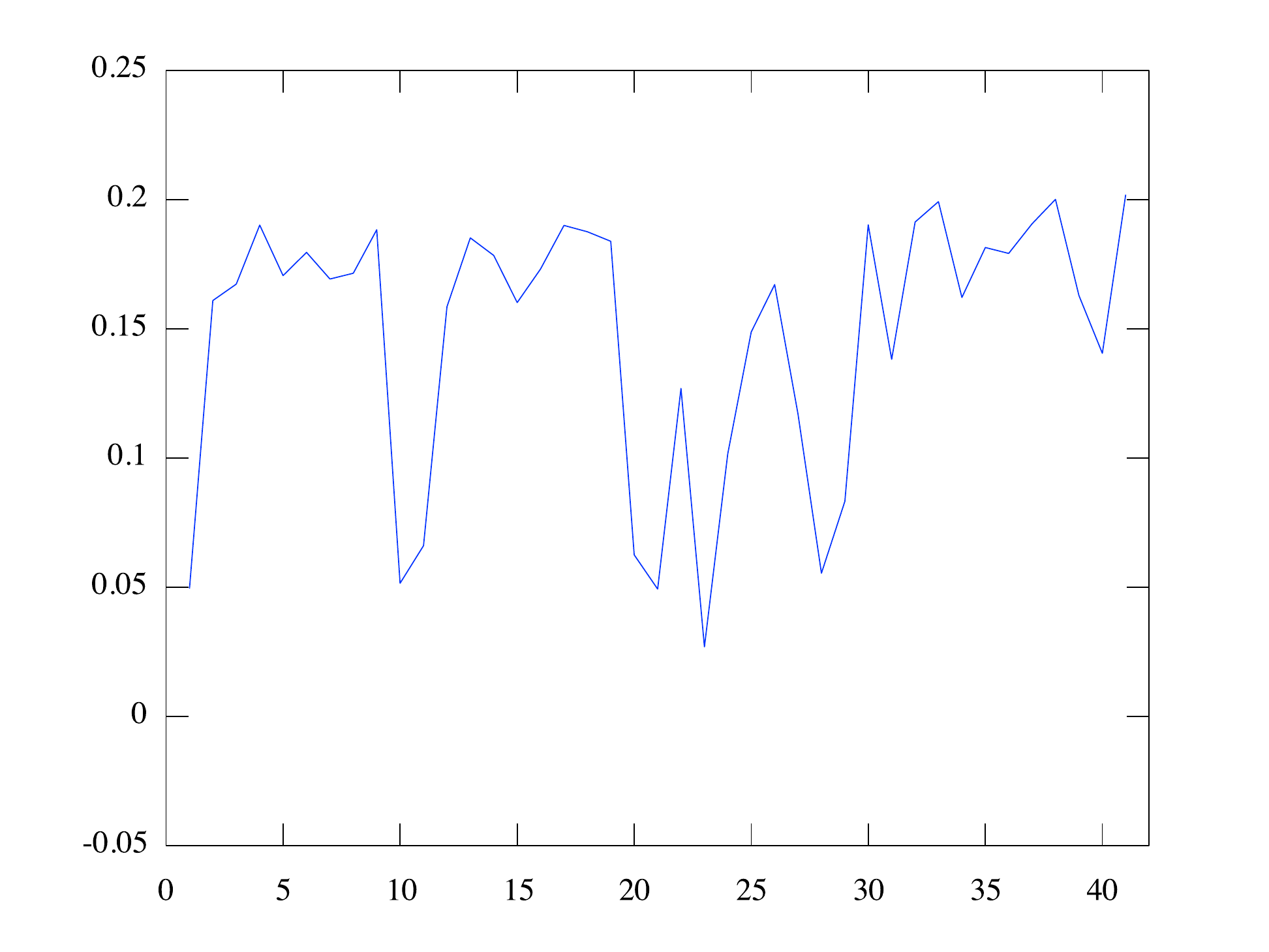}
	\includegraphics[angle=0,scale=0.3]{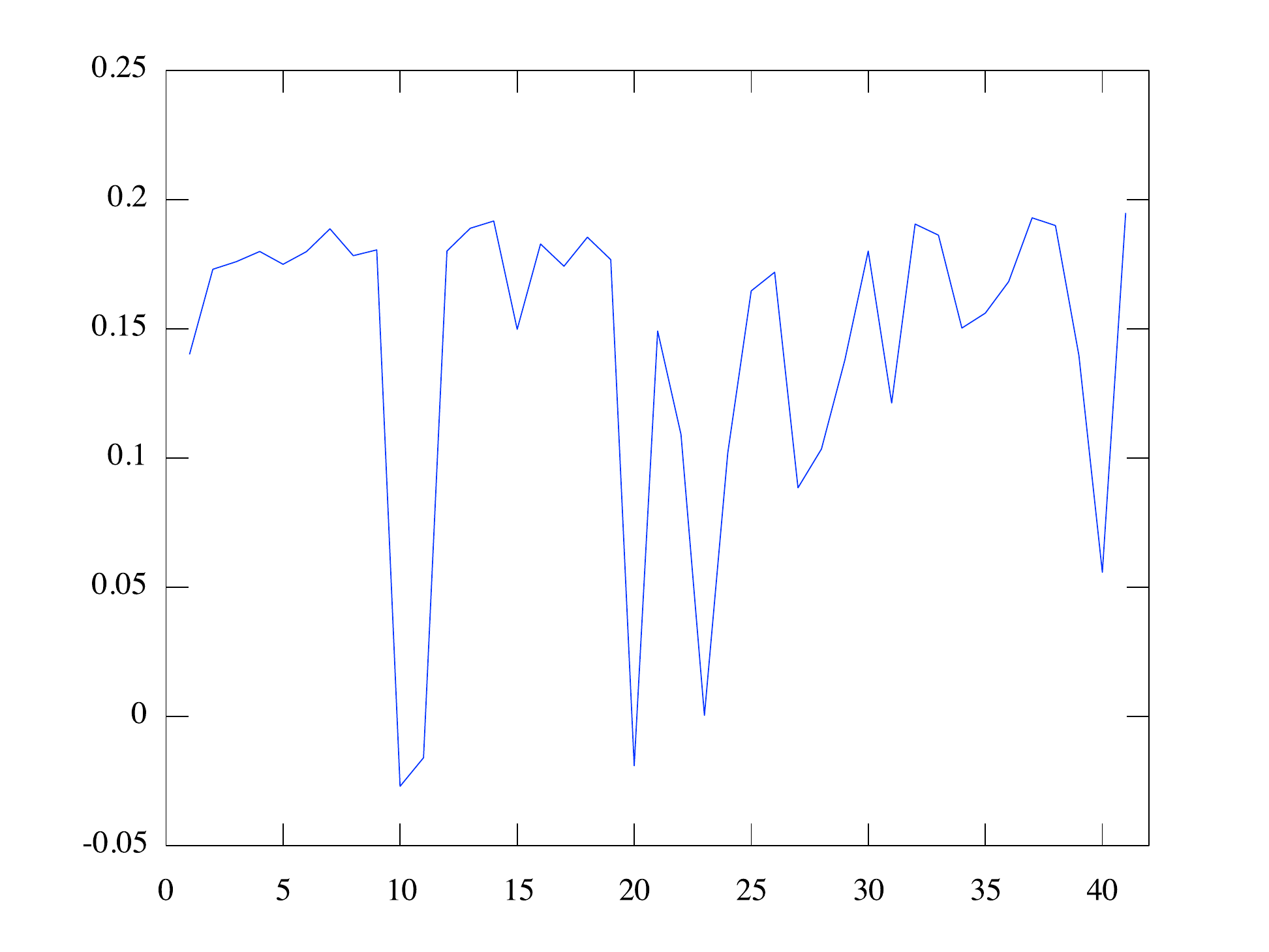}
	\caption{The first eigenvector of the Pearson correlation matrix. The index of the Brodmann areas are on the vertical axis and the value of the corresponding element of the eigenvector are on the horizontal axis. Left figure shows the eigenvector for the idle state, right one for tapping.}
	\label{eigv1tr_f}
\end{figure}

\begin{figure}
	\centering
	\includegraphics[angle=0,scale=0.08]{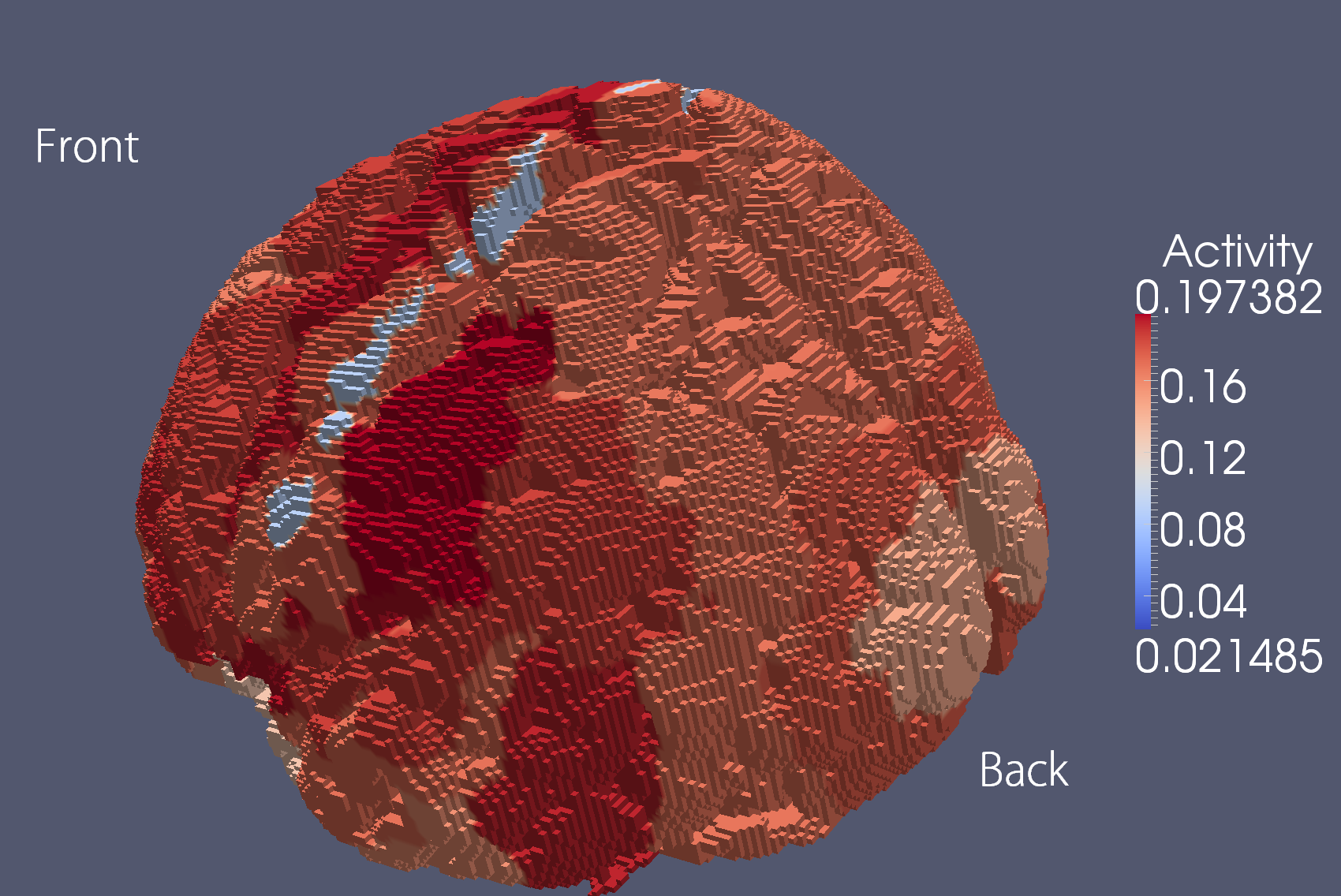}
	\includegraphics[angle=0,scale=0.08]{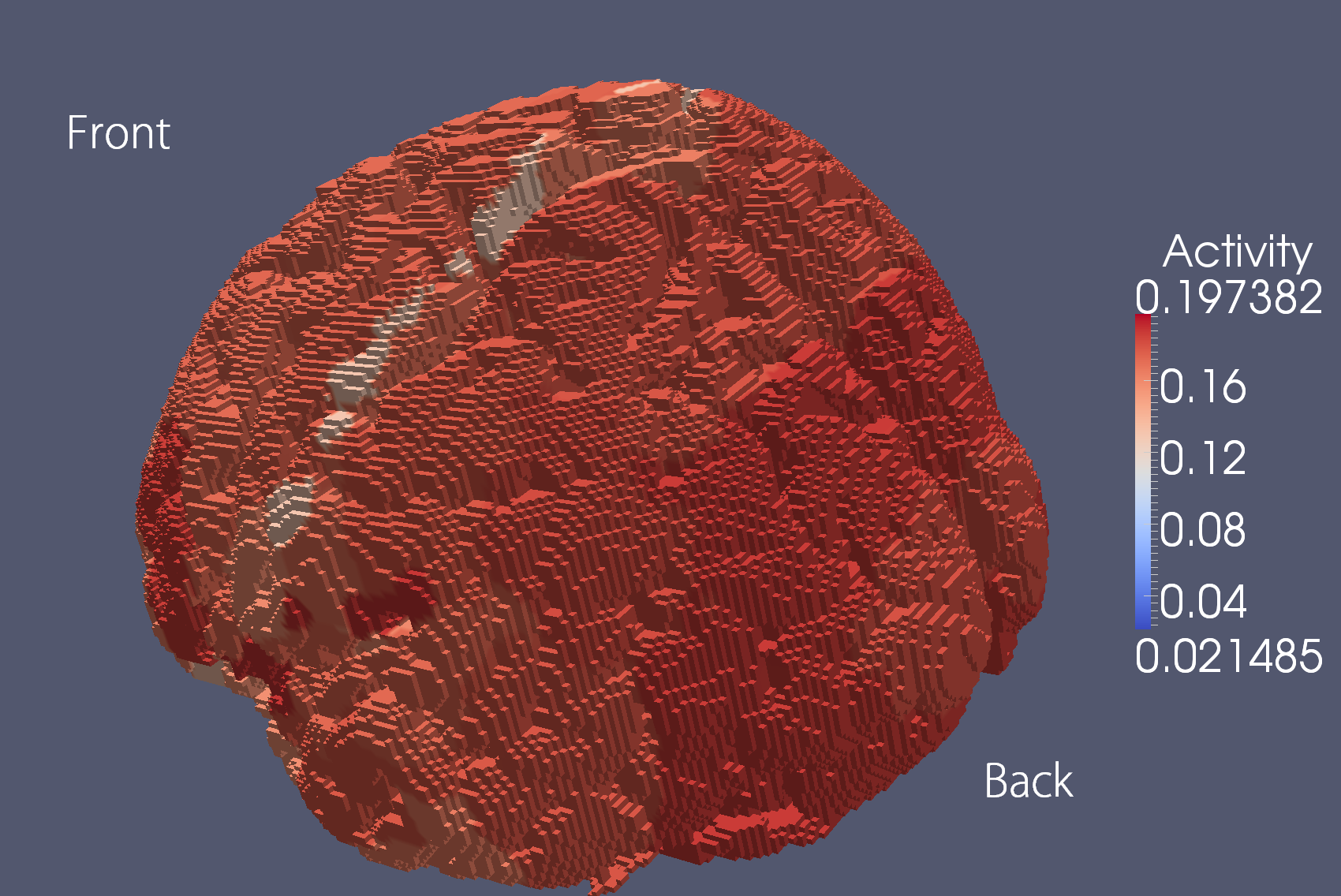}
	\caption{Visualization of the first eigenvector of the Pearson correlation matrix in the model brain. Values of the elements corresponding to different Brodmann areas are mapped into a colored scale. Left figure - idle state; right one - tapping.}
	\label{eigv1tr_f2}
\end{figure}

The participation ratio for the first eigenseries is equal to $34.393$ which is relatively close to the maximum value of $41$. This means that the first eigenseries collects the signal nearly evenly from almost all Brodmann regions. In the physiological sense this corresponds to a strong collective behavior in the brain. The participation ratio for this eigenseries is large independently on whether the brain is in the idle state or in the process of the "finger tapping" task. We have checked that by dividing the initial dataset into two subsets corresponding to "tapping" and "idle" states, respectively. The divided time series are shorter and thus rectangularity of the data and eigenvalues of $P$ matrix are a little different. Qualitatively however results are very similar.
The largest eigenvalue is equal to $24.407$ for tapping  $20.098$ for the idle state. They are indeed very close. We also check whether the corresponding eigenvectors are similar. The result is shown in Figure \ref{eigv1tr_f} as a plot and in Figure \ref{eigv1tr_f2} as an intensity map for Brodmann regions in the Talairach model of the brain. Already looking at the plot we see by inspection that the two plots in Figure \ref{eigv1tr_f} are similar. Both they have dips at the same positions even if differ a bit in the magnitude. If the underlying process had been related to the finger tapping, the two eigenvectors would have looked completely differently. This result provides us with an indication that eigenseries is rather related to a generic brain activity and not to any particular task-driven activity.

\subsection{Eigenseries corresponding to $\lambda_2$}
\begin{figure}
	\centering
	\includegraphics[angle=0,scale=0.3]{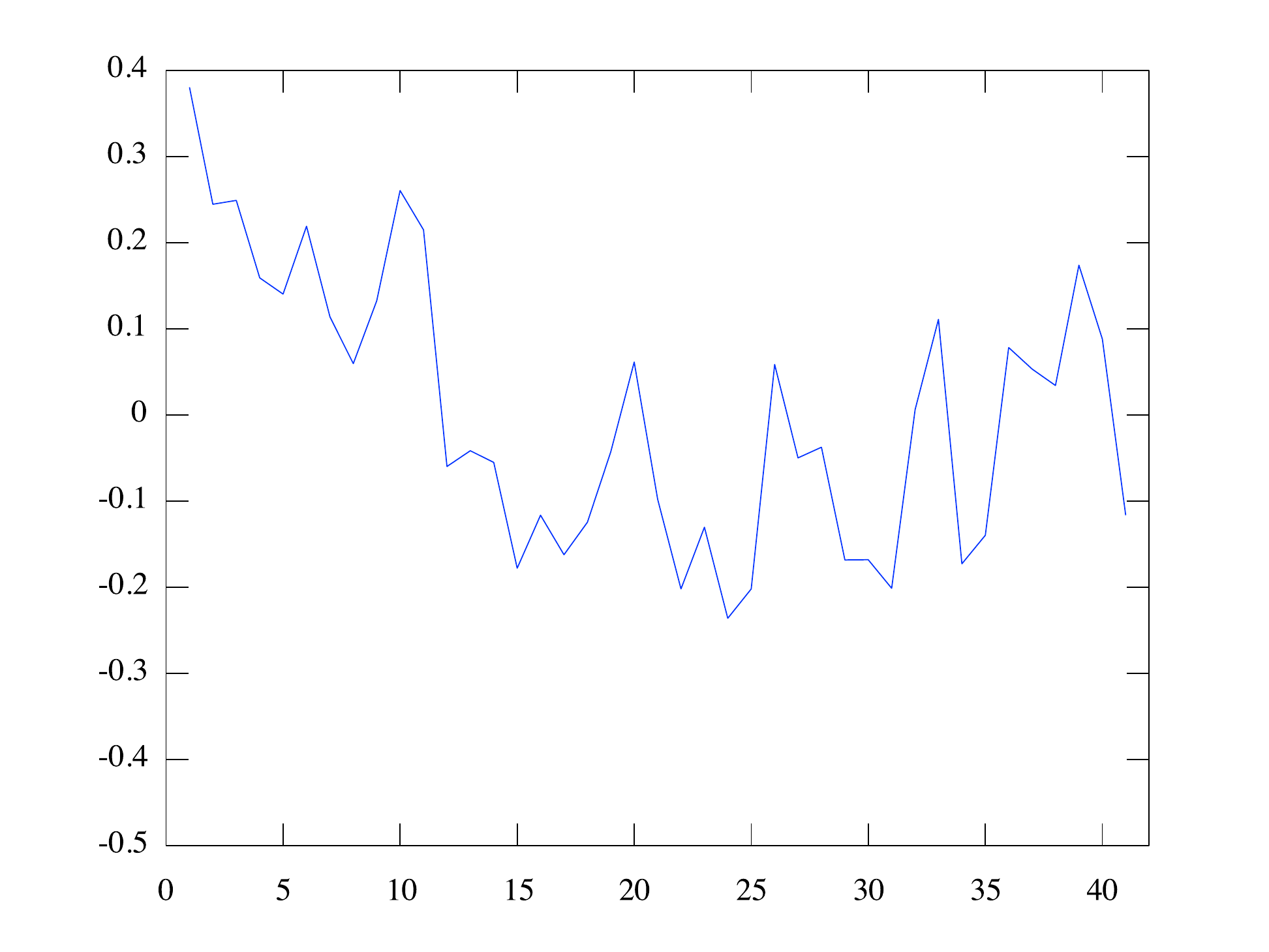}
	\includegraphics[angle=0,scale=0.3]{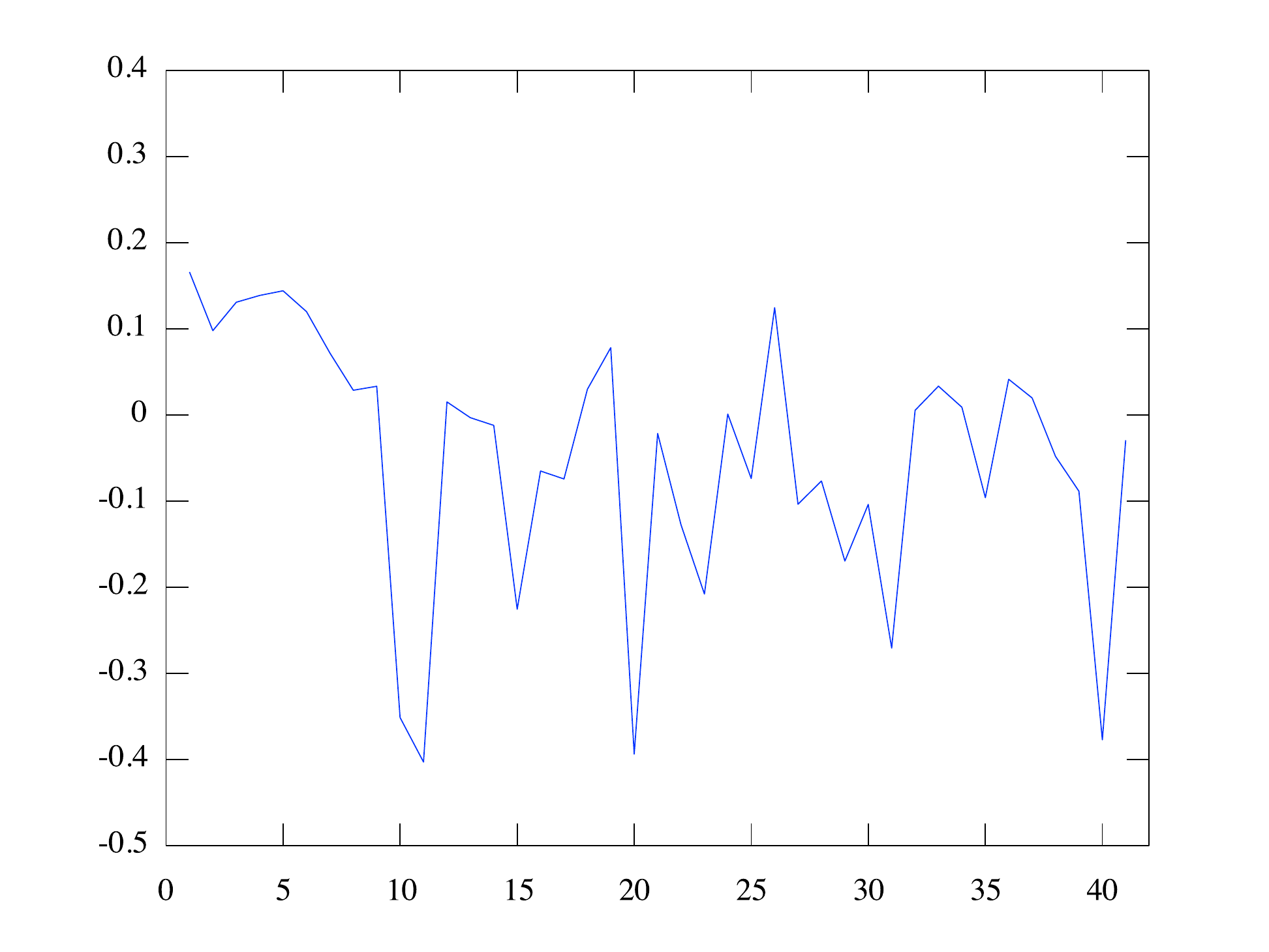}
		\caption{The second eigenvector of the Pearson correlation matrix. The index of the Brodmann areas are on the vertical axis and the value of the corresponding element of the eigenvector are on the horizontal axis. Left figure shows the eigenvector for the idle state and right one for tapping.}
	\label{eigv2tr_f}
\end{figure}

\begin{figure}
	\centering
	\includegraphics[angle=0,scale=0.08]{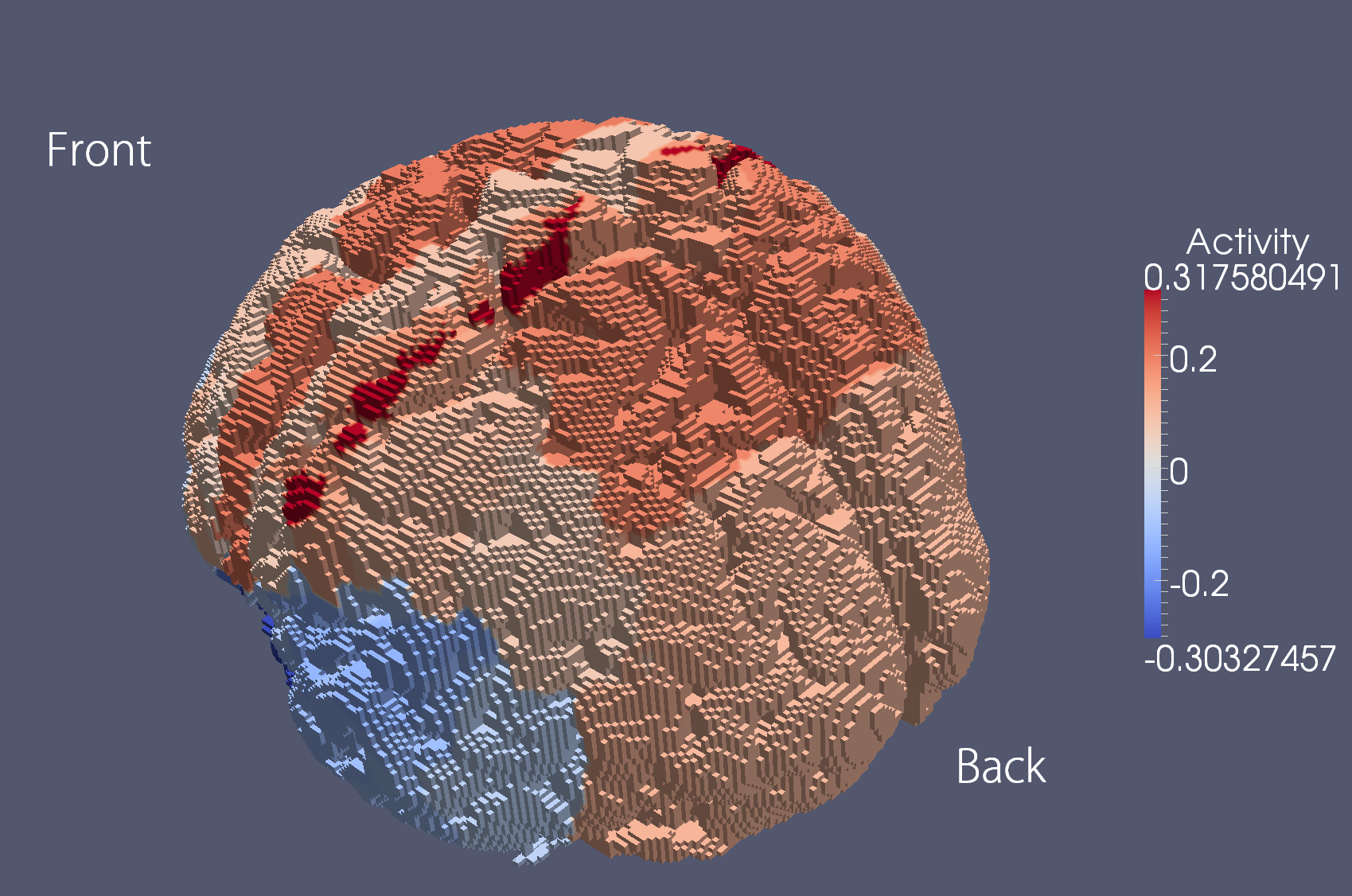}
	\includegraphics[angle=0,scale=0.08]{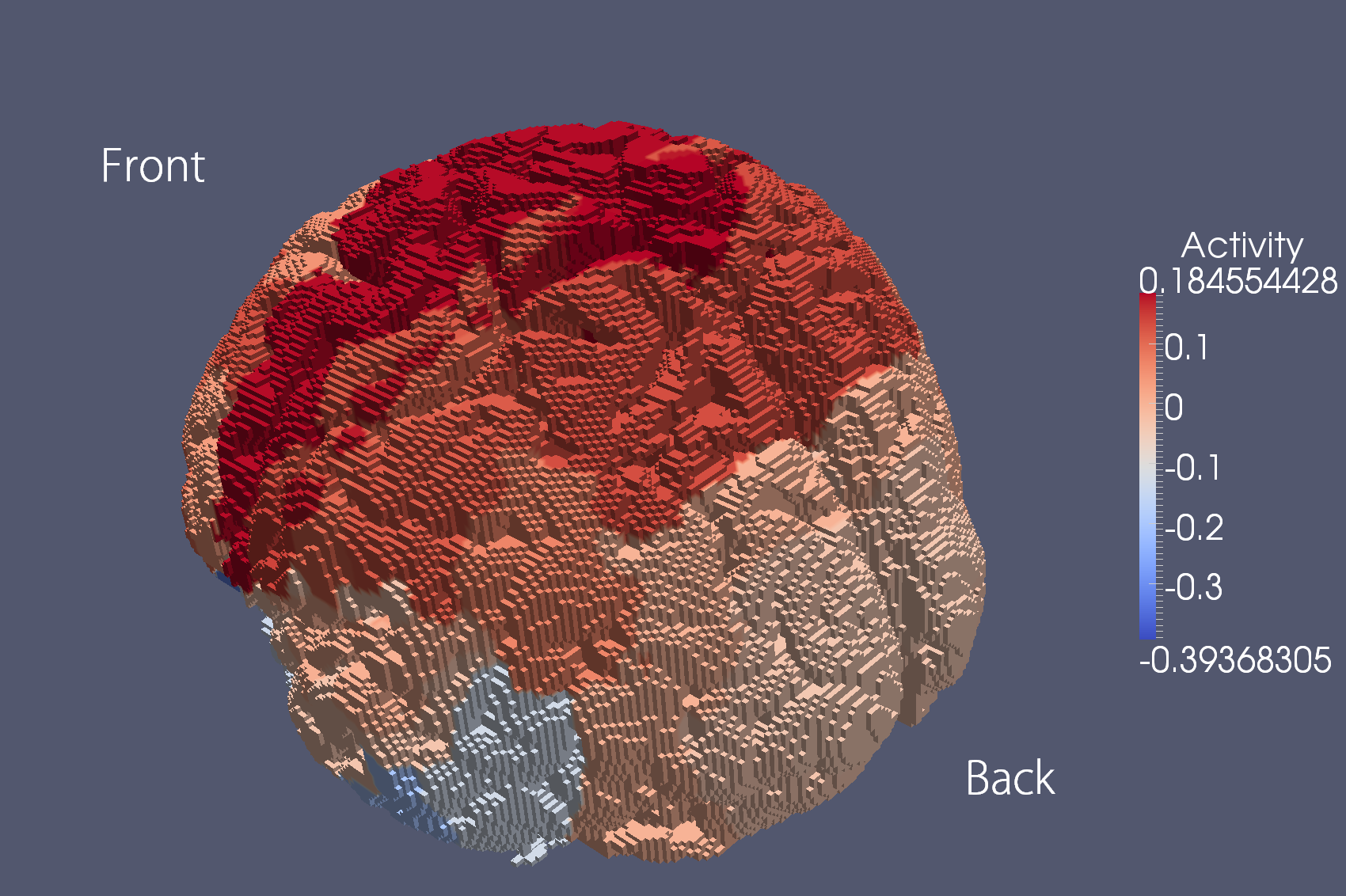}
		\caption{Visualization of the first eigenvector of the Pearson correlation matrix in the model brain. Values of the elements corresponding to different Brodmann areas are mapped into a colored scale. Left figure - idle state; Right figure - tapping.}
	\label{eigv2tr_f2}
\end{figure}

The second largest eigenvalue is equal to $4.478$ and thus it is greater than the cut-off $\lambda_+$. Again we expect that the information encoded in the eigenvector bears some statistically relevant information that exceeds the noise. The participation ratio for the eigenvector is equal to $16.42$. This indicates that less than a half of Brodmann areas take part in the corresponding brain activity. As before we divide our dataset into two separate subsets for the tapping and the idle state. The corresponding eigenvectors are presented as plots in Figure \ref{eigv2tr_f}. The horizontal axis shows the index of the Brodmann region and the vertical the value of the vector element that is proportional to its contribution to the collective behavior associated with this eigenvector (principal component). The vectors are also visualized in the brain model as colored Brodmann areas in Figure \ref{eigv2tr_f2}. Unlike for the first eigenvector, the differences here are significant between the vector for the tapping and the idle state and easy to see. The largest deviation can be seen in entries $9,10,20$ and $40$, which according to the table \ref{brodmann_numbers} correspond to Brodmann regions no. $9, 10, 25$ and $47$. These are parts of the prefrontal cortex ($9,10,25$) and frontal lobe ($47$). The former one is believed to be responsible for motor tasks planning. While interpreting the results one should remember that the values of the eigenvector elements are related to the contribution of the corresponding Brodmann region to the particular collective behavior represented by the given eigenseries and not to the activity of this region in the brain. 

Actually only the two leading eigenvalues pass the statistical significance criterion based on the cut-off $\lambda_+ = (1+\sqrt{r})^2$. Other eigenvalues are smaller than $\lambda_+$ and thus the associated with them principal components are basically shaped by statistical noise. Random Matrix Theory provides one with more advanced tools to obtain more rigorous criteria for the statistics of largest eigenvalues \cite{TW} close to the edge of the spectrum but the discussion of this issue lies beyond the scope of this paper.

\section{Conclusions} \label{s_C}

We have applied functional magnetic resonance imaging to study collective correlations of Brodmann areas in human cortex. The analysis was focused on the determination of collective correlations in which many Brodmann areas are simultaneously involved. To this end we used principal component analysis of the correlation matrix for fMRI intensities averaged over voxels in individual Brodmann regions. PCA analysis was supplemented by de-noising methods of random matrix theory. 

Our preliminary studies have shown that the principal component associated with the largest eigenvalue of the Pearson correlation matrix for fMRI signals is related to some physiological brain activity which is independent of whether the brain is involved in a task-driven activity or not. We have checked this by comparing the information content of the largest eigenvector in the idle period and the period of tapping. This comparison has shown that the eigenvector is basically statistically identical in the two periods. On the contrary, we have found that the principal component associated with the second largest eigenvalue is significantly different in the two states.

The principal components to the largest and the second largest eigenvalues differ also significantly in the number of Brodmann regions involved in the associated brain activities. This number has been estimated statistically by the participation ratio that in the former case has beed found to be of order of the number of studied regions while in the latter one to be less than a half of them. This means that in the first activity almost all Brodmann areas take part while in the second one only a selected subset. The analysis however was based on a single subject only, so it can be treated as a strong indication rather than the final result. We believe though that our preliminary studies reflect a generic pattern of correlations but it is to be verified on a larger group of subjects.

Using a random matrix criterion we have also shown that the remaining eigenvalues of the Pearson correlation matrix belong to the noisy part of the spectrum which means that the information content associated with the corresponding principal components is mostly dominated by the statistical noise as an effect of low statistics. Generally, in spectral analysis of multichannel data the signal-to-noise ratio is related to the aspect ratio of the data set that is calculated as the ratio of independent measurements to the number of data channels. This criterion is derived by the analysis of the benchmark for noise-driven data using random matrix theory. In neuroscience all typical experiments like fMRI, EEG or MEG are related to the analysis of multivariate data. We believe it is very useful to think in terms of random matrices and free probability while analyzing this type of data.

Using this methodology one can generalize the analysis to the case when in addition to correlations coming from collective behavior of many channels also temporal correlations are present in the data. In this case one can also study lagged correlations at different times by introducing the lagged correlations estimator $\frac{1}{T-1-\tau} \sum_{t=1}^{T-\tau} x_{it} x_{jt+\tau}$ which is a natural extension of Eq. (\ref{pc}) for the case of $\tau>0$.  Such correlation matrices are however generically non-Hermitian and require new developments in free probability. The first step in this direction has been done in \cite{bjn}.  

fMRI measurements have a good spatial resolution but rather poor temporal resolution.
On the contrary EEG measurements have a good temporal resolution and rather a poor spatial one. It would be tempting to attempt to combine the two methods in the future in a way allowing one to measure brain activity at localized brain regions with a better temporal resolution. This would improve the signal-to-noise ratio in the spectral analysis of the Pearson correlation matrix for Brodmann areas and increase the information content in a larger number of principal components. In effect one could extend the analysis beyond the components associated with the two largest eigenvalues.

\section*{Acknowledgements}
MAN and ZB acknowledge financial support by the Grant DEC-2011/02/A/ST1/00119 of the Polish National Centre of Science.



\end{document}